\documentclass[sigconf]{acmart}

\newcommand{\vectors}[1]{\lowercase{\mathbf{#1}}}

\usepackage{balance}

\usepackage{xcolor}
\usepackage{multirow}
\usepackage{microtype}
\usepackage{graphicx}
\usepackage{booktabs} % for professional tables

\usepackage{algorithm}
\usepackage{algorithmic}
\usepackage{amsmath}
\usepackage{amsfonts}
\usepackage{enumitem}
\usepackage{subcaption}
\AtBeginDocument{%
  \providecommand\BibTeX{{%
    \normalfont B\kern-0.5em{\scshape i\kern-0.25em b}\kern-0.8em\TeX}}}

\copyrightyear{2022}
\acmYear{2022}
\setcopyright{acmcopyright}
\acmConference[KDD '22] {Proceedings of the 28th ACM SIGKDD Conference on Knowledge Discovery and Data Mining}{August 14-18, 2022}{Washington DC Convention Center, USA.}
\acmBooktitle{Proceedings of the 28th ACM SIGKDD Conference on Knowledge Discovery and Data Mining (KDD '22), August 14-18, 2022, Washington, DC, USA}
\acmPrice{15.00}
\acmISBN{978-1-4503-XXXX-X/18/06}
\acmDOI{XXXXXXX.XXXXXXX}

\begin{document}
\fancyhead{}

% http://www.mlgworkshop.org/2022/

\title{Knowledge-aware Neural Collective Matrix Factorization for Cross-domain Recommendation}


%% work is done in my research intern in Baidu
% This work was performed when the first author was visiting Microsoft Research Asia
%% dawei corresponding author
% \author{PaperID:***}
\author{Li Zhang$^{1}$, Yan Ge $^{2}$, Jun Ma$^{3}$, Jianmo Ni$^{4}$ and  Haiping Lu$^{1}$}

\affiliation{
\institution{$^1$Department of Computer Science, University of Sheffield, Sheffield, United Kingdom} 
\institution{$^2$ Department of Computer Science, University of Bristol, Bristol, United Kingdom}
\institution{$^3$ Amazon Inc., Seattle, WA, USA} 
\institution{$^4$ Google Inc., USA}
\institution{$^{1}$\{lzhang72, h.lu\}@sheffield.ac.uk, $^2$yan.ge@bristol.ac.uk, $^3$junmaa@amazon.com, $^4$ jianmon.@google.com} 
\country{}
}

% \author{Li Zhang$^{1*}$, Lei Shi$^{2}$, Jiashu Zhao$^{3}$, Juan Yang$^{2}$, Tianshu Lyu$^{2}$, Dawei Yin$^{2\dagger}$ and  Haiping Lu$^{1}$}

% \affiliation{
% \institution{$^1$Department of Computer Science, University of Sheffield, Sheffield, United Kingdom} 
% \institution{$^2$Baidu Inc., Beijing, China}
% \institution{$^3$Department of Physics and Computer Science, Wilfrid Laurier University, Waterloo, Canada}
% \institution{$^{1}$\{lzhang72, h.lu\}@sheffield.ac.uk, $^2$\{shilei24, yangjuan03, lyutianshu\}@baidu.com, yindawei@acm.org} $^3$jzhao@wlu.ca
% \country{}
% }

\renewcommand{\shortauthors}{Li Zhang and Yan Ge, et al.}

\begin{abstract}
Cross-domain recommendation (CDR) can help customers find more satisfying items in different domains. Existing CDR models mainly use common users or mapping functions as bridges between domains but have very limited exploration in fully utilizing extra knowledge across domains. In this paper, we propose to incorporate the knowledge graph (KG) for CDR, which enables items in different domains to share knowledge. To this end, we first construct a new dataset \textit{AmazonKG4CDR} from the Freebase KG and a subset (two domain pairs: movies-music, movie-book) of Amazon Review Data. This new dataset facilitates linking knowledge to bridge within- and cross-domain items for CDR. Then we propose a new framework, KG-aware Neural Collective Matrix Factorization (KG-NeuCMF), leveraging KG to enrich item representations. It first learns item embeddings by graph convolutional autoencoder to capture both domain-specific and domain-general knowledge from adjacent and higher-order neighbours in the KG. Then, we maximize the  mutual information between item embeddings learned from the KG and user-item matrix to establish cross-domain relationships for better CDR. Finally, we conduct extensive experiments on the newly constructed dataset and demonstrate that our model significantly outperforms the best-performing baselines.

\end{abstract}


\maketitle

\section{Introduction}
\label{Introduction}

Cross-domain recommendation (CDR) \cite{fernandez2012cross} is a promising solution to the data sparsity problem in recommender systems. Conventional single-target CDR models leverage information from a richer (source) domain to improve the recommendation performance in a sparser (target) domain \cite{hu2013personalized, yuan2019darec, berkovsky2007cross}. To improve performance in both domains, recent dual-target CDR models \cite{man2017cross,zhu2019dtcdr,li2020ddtcdr} are proposed, which enables bidirectional transfer across domains with dual-learning mechanism \cite{zhang2019deep,he2016dual}.

Despite encouraging results from existing CDR models, several key issues remain unsolved \cite{zhu2021cross}. Firstly, current models, including the dual-target ones, can not simultaneously improve the performance in both source and target domains due to  negative transfer \cite{pan2009survey}. In general, the knowledge learned from the sparser domain is less accurate than that learned from the richer domain. Thus, the recommendation performance in the richer domain tends to decline if the transfer direction is simply inverted. Secondly, current CDR models mainly use common users \cite{man2017cross,zhu2019dtcdr} or mapping functions \cite{li2020ddtcdr} to build connections between domains. In real-life scenarios, relationships between items within or across domains can characterize item-wise semantic relatedness to help understand user-item interaction patterns \cite{wang2020exploiting}. However, current CDR models are inadequate in capturing such useful item-item relationships.

\begin{figure}[!t]
%	\vspace{0.5cm} 
	\centering
	\includegraphics[width=0.4\textwidth]{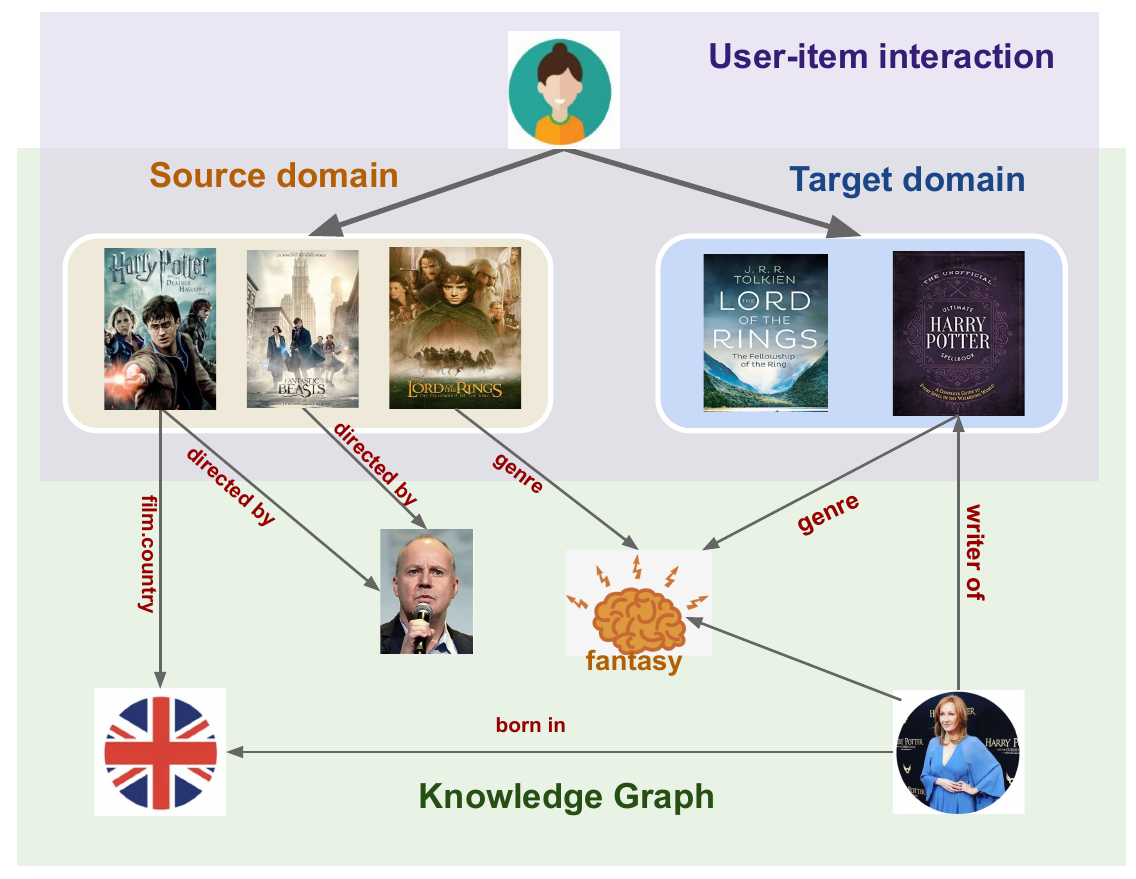}
	\caption{ \small Knowledge graph is a natural bridge that connects items from different domains. For example, ``\textit{Lord of the Ring}'' in movies can get connected with ``\textit{Harry Potter}'' in books via related genre \textit{Fantasy}. Such inter-domain knowledge can reveal similar semantic relations among items from different domains to further improve cross-domain recommendation. This paper constructs a new dataset and proposes a new model to achieve this goal.}%While current CDR models fail to capture such relations among items. }
	\label{fig:motivation}
\end{figure}

In this paper, we aim to address this gap by leveraging knowledge graph (KG), a natural bridge for items from different domains \cite{wang2017knowledge}. KGs can benefit the CDR task in multiple ways \cite{wang2018ripplenet}. First, rich and explicit connections among items in the KG can help improve the recommendation performance in each domain, particularly the sparser domain. As shown in Fig \ref{fig:motivation}, a user who has watched ``\textit{Harry Potter and the Deathly Hallows}'' is very likely to have interest in the movie ``\textit{Fantastic Beasts and Where to Find Them}'' (directed by the same director), which can be recommended with the assistance of domain-specific knowledge in the KG. Second, domains often share some domain-general information. For example, genre can characterize both book and movie domains. ``\textit{Lord of the Ring}'' (from movies), ``\textit{Harry Potter}'' (from books) can be closely connected in the KG via the related genre \textit{Fantasy}. KGs provide a natural bridge to build connections between domains. Leveraging  such information can help models understand target or source items by associating rich semantic relatedness among items from different domains and further improve recommendation performance.

To build KG-aware CDR, three unique technical challenges arise. (1) Though several datasets exist for KG-aware single-domain recommendation, no publicly-available dataset exists for KG-aware CDR. (2) To improve CDR, item (entity) embeddings (representations) learned from the KG should contain both domain-specific and domain-general information, which typically comes from different hops of neighbors in KG. The second challenge is to model both adjacent and higher-order relations in the item representation learning process.
(3) Item embeddings learned from the KG and those from the user-item interaction matrix should be closely related, e.g., highly correlated, so that cross-domain relationships can be effectively established. How to ensure this is the third challenge we need to overcome.

To address the challenges above, we construct a new dataset for KG-aware CDR and propose a novel KG-aware Neural Collective Matrix Factorization  (KG-aware NeuCMF) model. Firstly, we construct a new dataset named \textit{Amazon product Knowledge Graph for CDR} (\textit{AmazonKG4CDR)} using a subset (movie, book, and music) of the Amazon Review Data (2018) \cite{ni2019justifying} and the Freebase KG \cite{chah2017freebase,bollacker2008freebase}. Then, we propose a two-step framework for KG-aware CDR. 1) We train a shared autoencoder using a relational graph convolutional network (RGCN) on the knowledge graph following a contrastive learning-style \cite{kipf2016variational,schlichtkrull2018modeling}. GCN-based encoders learn a node's embedding by aggregating information from its neighbors via non-linear transformation and aggregation \cite{kipf2016semi}. Long-range node dependencies can be captured by stacking multiple GCN layers to propagate information for multiple hops \cite{pmlr-v80-xu18c}. This enables capturing both domain-specific and domain-general information from different hops of neighbors in the KG. 2) To establish cross-domain relationships, the embeddings learned from KG should be highly coherent with those from the user-item interaction matrix. Therefore, we incorporate the mutual information (MI) estimation \cite{belghazi2018mutual} into the neural collective matrix factorization (NeuCMF) framework. This mechanism allows our model to preserve both user-item interaction and KG information across items. Finally, we conduct extensive experiments on our newly constructed datasets and demonstrate that our model significantly outperforms the best-performing baselines, with up to 21\% (movie), 15.18\% (music) improvement, in terms of the mean absolute error (MAE) in movie-music domains recommendation.

% \begin{figure}[!t]
% %	\vspace{0.5cm} 
% 	\centering
% 	\includegraphics[width=0.4\textwidth]{fig/motivation.pdf}
% 	\caption{ \small Knowledge graph is a natural bridge that connects items from different domains. For example, ``\textit{Lord of the Ring}'' in movies can get connected with ``\textit{Harry Potter}'' in books via related genre \textit{Fantasy}. Such inter-domain knowledge can reveal similar semantic relations among items from different domains to further improve cross-domain recommendation. This paper constructs a new dataset and proposes a new model to achieve this goal.}%While current CDR models fail to capture such relations among items. }
% 	\label{fig:motivation}
% \end{figure}

%  21\%, 15.18\% in movie, music

In summary, our contributions are threefold:
\begin{itemize}
    \item We construct and leverage the knowledge graph for CDR task. To the best of our knowledge, this is the first time to apply KG information for CDR.
    \item  We propose a two-step KG-aware NeuCMF framework for KG-aware CDR, which enables learned item embeddings can capture both user-item interactions, domain-general, domain-specific information from the KG.
    \item We conduct extensive experiments on the newly constructed datasets. Experimental results show that our proposed model can significantly outperform most state-of-the-art CDR models.
    
\end{itemize}

 \begin{figure*}[!t]
%	\vspace{0.5cm} 
	\centering
	\includegraphics[width=0.9\textwidth]{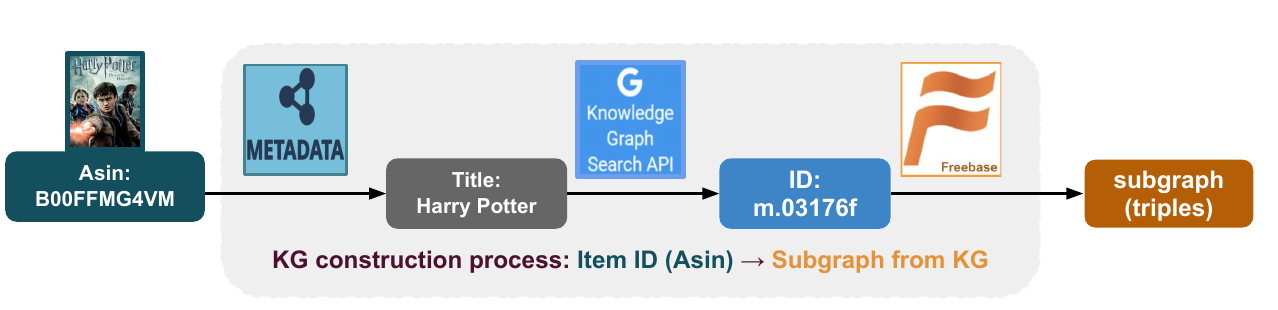}
	\caption{KG construction for Amazon products.}
	\label{fig:kg_construction}
\end{figure*}

\section{Related Work}
% \vspace*{-0.05in}
\subsection{Cross-Domain Recommendation}
% \vspace*{-0.05in}

Different from conventional single-domain recommendation, CDR can leverage information from source domain to improve the performance of target domain \cite{berkovsky2007cross,fernandez2012cross}, namely single-target CDR, which is a powerful tool to deal with the data sparsity problem. These approaches extend the single-domain recommendation models by utilizing same contents, such as tags, reviews \cite{fernandez2014exploiting,Yuan2019DARecDD}, common items or users \cite{singh2008rel,hu2018conet,lian2017cccfnet} as the bridge between and transfer information between domains \cite{hu2013personalized,Loni2014CrossDomainCF,Sahebi2015ItTT,Sahebi2014ContentBasedCR}.

The single-target CDR approaches only focus on how to leverage the source domain to help improve the recommendation accuracy on the target one, but not vice versa. Recently, dual-target CDR mothods \cite{man2017cross,zhu2019dtcdr,li2020ddtcdr} has been proposed to improve the performance on both source and target domains simultaneously by leveraging dual-transfer learning strategies \cite{zhang2019deep,he2016dual}. However, as referred to as Negative Transfer \cite{pan2009survey}, this idea does not work, because the knowledge learned from the sparser domain is less accurate than that learned from the richer domain, thus the recommendation accuracy on the richer domain is more likely to decline by simply
and directly changing the transfer direction. Therefore, dual target CDR demands novel and effective solutions. None of the current CDR models can indeed improve the performance on both domains simultaneously, and they are significantly hindered  by  limited  information  and  connections between two domains.

\subsection{Knowledge Graph for Recommendation}

In recent years, introducing recommendations with the KG as side information has attracted considerable interest \cite{wang2018ripplenet,wang2017knowledge,KGAT19}. A KG is a heterogeneous graph, where nodes represent as entities, edges represent relations between entities and a fact in KG is usually represented in the form of a triple (\textit{head entity}, \textit{relation}, \textit{tail entity}) \cite{wang2017knowledge}. KGs contain rich semantic relatedness among items and incorporating KGs in RS can help explore the latent connections and provide explanations for recommended items \cite{guo2020survey}. Currently, KG-aware RS models are only for the single-domain RS \cite{catherine2016personalized,wang2018ripplenet,tang2019akupm,zhao2019intentgc,wang2017knowledge,KGAT19}. While one bottleneck for CDR is lacking of connections between domains. since KGs can naturally connect different domains, it would be promising by incorporating KG in the user-item interaction matrix for better cross-domain recommendation performance.

%  \begin{figure*}[!t]
% %	\vspace{0.5cm} 
% 	\centering
% 	\includegraphics[width=0.9\textwidth]{fig/kg_construct2.pdf}
% 	\caption{KG construction for Amazon products.}
% 	\label{fig:kg_construction}
% \end{figure*}

\section{KG-aware NeuCMF models}
In this section, we present the technical details of our proposed CDR model, KG-aware Neural CMF (KG-NeuCMF) that aims to improve the performance of CDR by leveraging the KG. This section first introduces how to construct the knowledge graph for items. Then we formulate the task and present our proposed framework: KG-NeuCMF.

\subsection{KG Construction for CDR}
\label{sec:kg construction}

To develop a knowledge-aware CDR system, a key issue is how to obtain rich and structured knowledge information for items. Existing research works use
%data sets or methods either use
 side information from the original recommender system, such as tags and reviews. We argue that the KG information will provide additional useful information to the CDR task, since the intra-domain relationship among items can be captured.
 % such as tag, review containing limited useful information. 
 In this paper, we present \textit{AmazonKG4CDR V1.0}, a new dataset linking KG information for CDR, which can be useful for researchers in the related areas to explore possible approaches with the rich KG information. 

We use the widely used dataset, Amazon Review Data (2018) \cite{ni2019justifying}, covering various domains, from which we select a subset that includes two domain pairs: movie-music, movie-book, which are being linked together through a common user ID identifying the same user. On the KG side, we use the well-known KG: Freebase \cite{bollacker2008freebase}. It stores facts by triples of the form $<head> <relation> <tail>$. Since Freebase shut down its services, we use its latest public version. We map items into Freebase entities via title matching if there is a mapping available.  Fig.\ref{fig:kg_construction} shows the whole linkage process. Since we only have item Asins (IDs of Amazon products), we need to get items’ titles from the Amazon Review metadata first\footnote{https://nijianmo.github.io/amazon/index.html}. These titles are later used to get KG entity IDs from \textit{The Knowledge Graph Search API}, which are used to extract the graph information from Freebase.
% We take triplets that involve two-hop neighbor entities of items into consideration.

During the linkage process, we have dealt with several problems that will affect the quality of the extract knowledge graph. First, the correctness of the extracted KG entity IDs should be ensured. For example, a query is \textit{``Harry Potter''} (a book name), and returned results can be both movies and books.
So, we filter returned results by their type and name to ensure extracted IDs are correct.
% In order to ensure the extracted IDs from \textit{The Knowledge Graph Search API} are correct ( Movies may have the same title with books), we use both returned query type and name to verify the correctness of the IDs.
 To ensure the KG quality, we preprocess the extracted KG by filtering out infrequent entities (e.g., lower than 10 in both datasets) and retaining the relations appearing in at least 100 triplets.

\subsection{Problem Statement}
In this paper, we study the problem of KG-aware CDR. Formally, we are given two domains, a source domain $\mathcal{S}$ (e.g., movie recommendation) and a target domain $\mathcal{T}$ (e.g., book recommendation) that can be represented as two user-item interaction matrices $\mathbf{R}_{\mathcal{S}}$ and $\mathbf{R}_{\mathcal{T}}$, where $r_{ui} = 1$ indicates that user $u$ engages with item $i$, otherwise $r_{ui}=0$. In real online shopping platforms (e.g., Amazon), users in domain $\mathcal{S}$ and domain $\mathcal{T}$ often overlap, meaning that they have purchased items in both domains. The set of users in both domains are shared, denoted by $\mathcal{U}$ (of size $m$ = $\left\vert\mathcal{U}\right\vert$). In our setting, there is no overlap of items between two domains and each item only belongs to one single domain. Denote the set of items in $\mathcal{S}$ and $\mathcal{T}$ by $\mathcal{I_\mathcal{S}}$ and $\mathcal{I_T}$ with size $n_\mathcal{S}$ = $\left\vert\mathcal{I_{S}}\right\vert$) and $n_\mathcal{T}$ = $\left\vert\mathcal{I_{T}}\right\vert$ respectively. Additionally, we also have a knowledge graph $\mathcal{G}$, a multi-relational graph, containing rich facts about items. Each fact in the KG is represented as a triple (head entity,relation,tail entity) (($h,r,t$)) \cite{wang2017knowledge}. The KG can represent large-scale information from multiple domains \cite{ehrlinger2016towards}. In recommendation scenarios, an item in the user-item interaction matrix corresponds to an entity in the KG.

Given $\mathbf{R}_{\mathcal{S}}$ and $\mathbf{R}_{\mathcal{T}}$ as well as the knowledge graph $\mathcal{G}$, we aim to predict whether user $u$ will engage with item $i$ with which the user has no interaction before. Our goal is to learn a prediction function 
$\hat{y}_{ui}$ = $f$ $(u,i\mid \Theta, \mathbf{R}_{\mathcal{S}},\mathbf{R}_{\mathcal{T}},\mathcal{G})$, where $\hat{y}_{ui}$ denotes the probability (or the rating score) that user $u$ will engage with item $i$ and $\Theta$ denotes the model parameters of function $f$.

\subsection{Methodology}

 \begin{figure*}[!t]
%	\vspace{0.5cm} 
	\centering
	\includegraphics[width=0.99\textwidth]{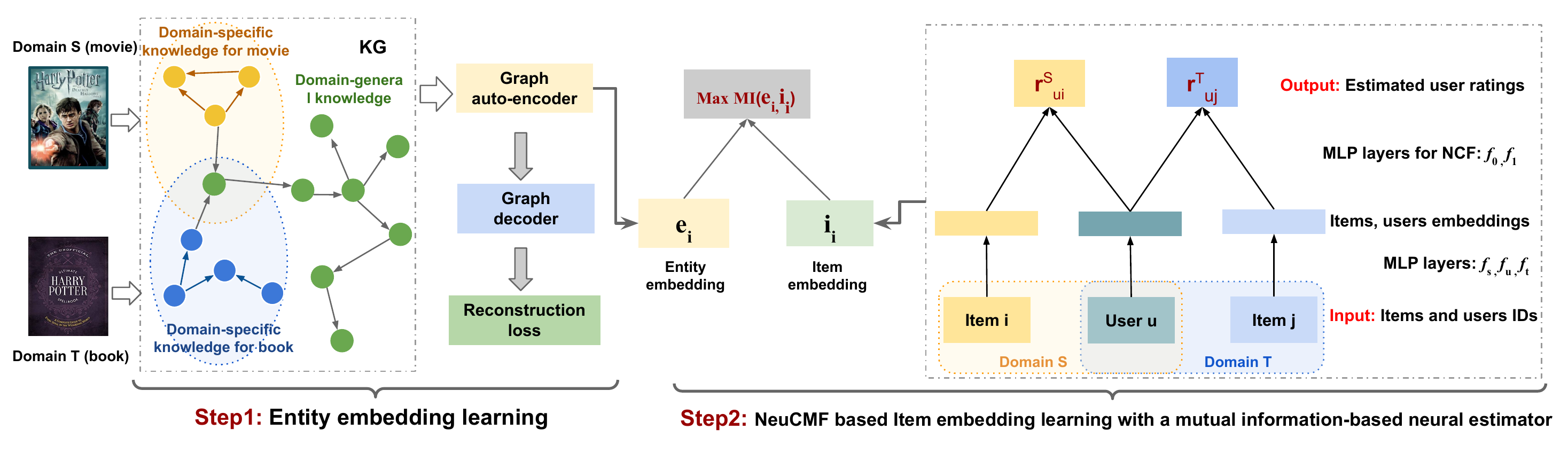}
	\caption{\small The framework of our model: KG-aware NeuCMF. It learns item representations from both KG (left) and user-item interaction matrices (right). Entity (item) representations learned from KG contain both domain-specific and domain-general information by utilizing graph autoencoding strategy, which can help assist the CDR task. Item embeddings are learned by a neural CMF model. To ensure the two types of embeddings are highly correlated, we maximize their MI by the neural mutual information estimator (middle).}
	\label{fig:framework}
\end{figure*}

In this subsection, we present the technical details of our proposed model, KG-aware Neural CMF (KG-NeuCMF) that aims to improve the performance of CDR by leveraging the KG. Fig.\ref{fig:framework} shows the overview of the proposed framework. In the first stage, we propose to learn KG-level representations by exploiting a multi-layer RGCN \cite{schlichtkrull2018modeling} through the encode-decode paradigm by minimizing the reconstruction loss that follows a contrastive learning-style convention \cite{gvae}. This step aims to learn item embeddings containing both domain-specific and domain-general information from different hops of neighbors in KG. In the second-stage, we learn item and user embeddings by borrowing ideas from the CMF framework \cite{singh2008rel} and neural CF (NCF) \cite{he2017neural}. Instead of jointly factorizing the two user-item interaction matrices directly as in CMF, we propose to utilize neural networks to jointly learn the two matrices by sharing user latent representations. Finally, item representations learned from KG and  user-item interaction matrix should be highly correlated. To quantify such correlation, we also exploit to maximize MI \cite{belghazi2018mutual} between the two types of representations.

\subsubsection{Entity embedding learning}
To utilize the KG in our task, we first need to learn entity representations.  We do this by training a graph autoencoder model in the unsupervised fashion and learn representations in an encode-decode paradigm \cite{gvae,schlichtkrull2018modeling}. We employ RGCN \cite{schlichtkrull2018modeling} as our encoder that learns an entity embedding by aggregating information from its adjacent neighbors via non-linear transformation and aggregation dependent on the connecting relation, which can be denoted as 

%  $\hat{\vectors{e}}_{{i}}^{(l)}$

\begin{equation} 
f_{en}(\vectors{e}^{(l)}_{{i}},\vectors{e}^{(l)}_{{j}})  = \sigma (\mathbf{W}^{(l)}_{0} \vectors{e}^{(l)}_{{i}} + \sum_{r \in \mathcal{R}}\sum_{{j} \in \mathcal{N}_{i}^{r}} \dfrac{1}{c_{ij}} \mathbf{W}^{(l)}_{r} \vectors{e}^{(l)}_{{j}}),
\label{eq:kg_encoder}
\end{equation}
where $\vectors{e}^{(l)}_{{i}},$ $\vectors{e}^{(l)}_{{j}}$ are the hidden state of node $i$ and node $j$ in the $l-$th layer of the encoder, $\sigma$ is an activation function such as ReLU, $\mathbf{W}^{(l)}_{0}$, $\mathbf{W}^{(l)}_{r}$ are (learnable parameters) relation-specific transformation mapping matrices depending on the type of edge, $c_{ij}$ is problem-specific normalization constant that can either be learned or chosen in advance, and $\mathcal{N}_{i}^{r}$ denotes the set of neighbors of node $i$ under relation $r \in \mathcal{R}$. Through this operation, the local proximity structure and related semantic information can be successfully captured and stored in the new representation of each entity.  Long-range node dependencies can be captured by stacking multiple graph encoder layers and this mechanism ensures that distinct domains can be connected via the information propagation. 

The decoder can be any scoring function of KG embedding methods \cite{wang2017knowledge} that are used to measure the plausibility of each fact ($h,r,t$). Following \cite{schlichtkrull2018modeling}, we use DisMult \cite{yang2014embedding} factorization as the scoring function, which is well known for its simplicity and efficiency and a triple ($h,r,t$) is scored as  

\begin{equation} 
f_{de}(\vectors{e}_{{h}}, \vectors{r},\vectors{e}_{{t}}) = \vectors{e}_{{h}} \mathbf{R}_{r} \vectors{e}_{{t}},
\label{eq:kg_decoder}
\end{equation}
where $\vectors{e}_{{h}},\vectors{e}_{{t}} \in \mathbb{R}^{d} $ are  encoded features vector for entity $h$ and $t$, and each relation $\vectors{r}$ is associated with a diagonal matrix $\mathbf{R}_{r}\in \mathbb{R}^{d\times d}$. 

We train the encoder and decoder with negative sampling. We construct an equal number of negative samples by randomly replacing the head entity or tail entity of each positive sample and the overall set of samples are denoted by $\mathcal{M}$. Then we minimize the cross-entropy loss of positive and negative node pairs

\begin{equation}
    \mathcal{L} = \sum_{(\vectors{e}_{{h}},\vectors{r}, \vectors{e}_{{t}} ,y)\in \mathcal{M}} (y log f_{de}(\vectors{e}_{{h}},\vectors{r},\vectors{e}_{{t}}))) + (1-y) log (1-f_{de}(\vectors{e}_{{h}},\vectors{r},\vectors{e}_{{t}})).
\end{equation}

\subsubsection{NeuCMF module}

Typically the user-item interaction matrices are highly sparse and it is beneficial to learn them simultaneously \cite{singh2008rel}. Collective matrix factorization (CMF) jointly factorizes two matrices by sharing the user latent factors. Motivated by neural CF (NCF) \cite{he2017neural}, we propose to utilize neural networks to jointly learn the two matrices by sharing user latent representations as shown in Fig. \ref{fig:framework}.
The predicted scores in two domains are 
\begin{equation}
    r_{ui}^{\mathcal{S}}= f_{0}(f_{u}(\vectors{u}_{u}), f_{s} (\vectors{i}_{i}^{\mathcal{S}}),
\end{equation}
\begin{equation}
    r_{uj}^{\mathcal{T}}= f_{1}(f_{u}(\vectors{u}_{u})), f_{t} (\vectors{i}_{j}^{\mathcal{T}})),
\end{equation}
where $\vectors{u}_{u}$, $\vectors{i}_{i}^{\mathcal{S}}$ and $\vectors{i}_{j}^{\mathcal{T}}$ are represented one-hot vectors of users, items from domain $\mathcal{S}$ and domain $\mathcal{T}$ respectively. Only the element corresponding to that index is 1 and all others are 0. $f_{u}$, $f_{s}$ and $f_{t}$ can be  multi-layer perceptron (MLP) that project sparse representations to dense vectors. The obtained embeddings are then feed into two separate multi-layer neural architectures to map the latent vectors to predict scores $r_{us}^{\mathcal{S}}$, $r_{ut}^{\mathcal{T}}$ for the two domains. Given $\mathbf{R}_{\mathcal{S}}$ and $\mathbf{R}_{\mathcal{T}}$, we minimize the two reconstruction losses $\mathcal{L}_{\mathcal{S}}$ and $\mathcal{L}_{\mathcal{T}}$ with the predicted scores.

The NeuCMF module connects two domains only by the common users, and fails to capture the relations among items. The item embedding learned from KG can capture both domain-specific and domain-general knowledge, thus will be effective for both single-domain and cross-domain recommendation. Intuitively, the learned item embedding from user-item interaction matrices should be highly correlated to the KG-level embeddings. Therefore, this motivates us to exploit to maximize MI \cite{belghazi2018mutual} between the two types of representations to guarantee their highly correlated relationship. We design our neural mutual information estimator based on a discriminator $\mathcal{D}(x,y)$ for their pairwise relationships, to provide probability scores for sampled pairs. To be specific, we generate positive samples as ($\vectors{e}_{i}$,$\vectors{i}_{i}$) ($\vectors{i}$ can come from domain $\mathcal{S}$ and domain $\mathcal{T}$, half-half) and negative samples are generated by associating sampled items with fake embeddings based on shuffling strategy \cite{velickovic2019deep}. We define the loss function as:
\begin{equation}
    \mathcal{L}_{mul} = -\frac{1}{N} (\sum_{i=1}^{N_{pos}} \mu (\vectors{i}_{i},\vectors{e}_{i}) log\sigma(\vectors{i}_{i},\vectors{e}_{i}) +
    \sum_{i=1}^{N_{neg}} \mu (\tilde{\vectors{i}_{i}},\vectors{e}_{i}) log \sigma( \tilde{\vectors{i}_{i}},\vectors{e}_{i})),
\end{equation}
where $N = N_{pos} + N_{neg}$, $N_{pos}, N_{neg}$ denotes the number of positive and negative samples, $\mu(\cdot)$ is an indicator function, $\sum_{i=1}^{N_{pos}} \mu (\vectors{i}_{i},\vectors{e}_{i})=1$ and  $ \sum_{i=1}^{N_{neg}} \mu (\tilde{\vectors{i}_{i}},\vectors{e}_{i})=1$ corresponds to positive and negative pair samples. We aim to
minimize $\mathcal{L}_{mul}$, which is equivalent to maximize the mutual information, to jointly preserve the KG-level and user-item interaction information.

The final loss includes: the loss ($\mathcal{L}_{\mathcal{S}}$) of source and loss ($\mathcal{L}_{\mathcal{T}}$) of target recommendation with the mutual information maximization loss $\mathcal{L}_{mul}$. 
The objective is to minimize the overall loss $\mathcal{L}$ as follows:
\begin{equation}
\mathcal{L} = \mathcal{L}_{\mathcal{S}}(\Theta_{\mathcal{S}}) + \mathcal{L}_{\mathcal{T}}(\Theta_{\mathcal{T}}) + \mathcal{L}_{mul}(\Theta_{mul}) + \lambda \lVert \Theta \rVert,
\end{equation}
where $\Theta$ = $\Theta_{\mathcal{S}}$ $\cup$ $\Theta_{\mathcal{T}}$ $\cup$ $\Theta_{\mathcal{L}_{mul}}$. Note that $\Theta_{\mathcal{S}}$ and $\Theta_{\mathcal{T}}$ share user embeddings. The objective function can be optimized by stochastic gradient descent (SGD) and its variants like adaptive moment method (Adam) \cite{kingma2014adam}.

\section{Experiment}
\label{sec:experiment}

\subsection{Dataset}
We use the Amazon Review Data (2018) \cite{ni2019justifying} that is widely used for product recommendation. It contains users' rate (ranging from 1 to 5) for product from various domains. We select a subset that includes two domain pairs: movie-music(MM), movie-book(MB), which are being linked together through a common user ID identifying the same user. We construct the knowledge graph for each item by utilizing Freebase and take triplets that involve two-hop neighbor entities of items into consideration. The basic statistics details are presented in Table 1. The recommendation task can be formulate as the regression (rating) or the binary classification (recommend or not) tasks. Following \cite{ricci2011introduction}, we evaluate the recommendation performance based MAE, F1\_score (Threshold of positive rating is 4) for the regression and classification performance, respectively.

\subsection{Baselines}
\label{sec:baselines}

 To validate the performance of the proposed model, we compare the performance with five representative models, in which two single-domain RS models (MF, NCF) and three CDR models (CMF, CoNet, DDTCDR) using the publicly released implementations. 
\begin{itemize}
    \item \textbf{MF} \cite{koren2009matrix}. Matrix Factorization (MF) is a classic latent
factors CF approach which learns the user and item factors via matrix factorization in each domain separately.
    \item \textbf{NCF} \cite{he2017neural}. Neural Collaborative Filtering (NCF) is a neural network architecture to model latent features of users and items using CF method. The NCF models are trained separately for each domain without transferring any information.
    \item \textbf{CMF} \cite{singh2008rel}. Collective Matrix Factorization (CMF) jointly factorizes matrices of each domains. In our scenarios, The shared user factors enable knowledge transfer between cross domains .
    \item \textbf{CoNet} \cite{hu2018conet}. Collaborative Cross Networks (CoNet) enables dual knowledge transfer across domains by introducing cross connections from one base network to another and vice versa.
    \item \textbf{DDTCDR} \cite{li2020ddtcdr}. Deep Dual Transfer Cross Domain Recommendation (DDTCDR) learns latent orthogonal mappings across domains and provides cross domain recommendations by leveraging user preferences from all domains.
\end{itemize}

\begin{table}[!t]
\centering
\caption{ Statistics of the dataset.}
\begin{tabular}{l|cc|cc}
\toprule
& \multicolumn{2}{c}{Domain: Music-Movie} & \multicolumn{2}{|c}{Domain: Book-Movie} \\ \cline{2-5}
 & Music    & Movie     & Book    & Movie  \\ \toprule
Users  & 4,196   & 4,196   & 3,977 & 3,977 \\ 
Items  & 7,412   & 10,919   & 11,372 & 8,118 \\
Interactions & 21,986 & 49,027  & 22,214 & 29,245 \\ 
\midrule
% Sparsity & 21,986 & 49,027  & 22,214 & 29,245 \\ \midrule
% Com\_kgid    & 6,372  & 8,803 & 10,355& 6,611 \\
Entities & 85,612  & 387,178  & 258,999  & 990,141\\
Relations  & 155  & 340 & 127& 295   \\
Triples& 288,731  & 610,314  & 522,814  & 1,787,190 \\ \bottomrule 
\end{tabular}
\end{table}

\subsection{Implementation details}
\label{sec:hyper-details}
In the KG-pretrain step, we utilize a two-layer RGCN 
as the encoder to obtain entity embeddings. In the NeuCMF module, we apply one-layer neural networks to project the one-hot vectors of users, and items to low-dimensional embedding vectors and $f_{0}$ and $f_{1}$ are two one-layer neural networks to map the latent vectors to predict scores. Throughout the experiments, the embedding size is tuned in the range of [8,16,32] and we use the Adam optimizer \cite{kingma2014adam} with learning rate 0.001, L2 regularization 0.0001. For each dataset, the ratio of training, evaluation, and test set is 6 : 2 : 2 \cite{wang2019knowledge}. We employ the early stopping strategy based on the validation accuracy with a
window size of 10 (we will stop training if the validation loss does not decrease for 10 consecutive epochs) and train 200 epochs at most. We report results over 20 runs with random weight matrix initialization. For a fair comparison, we set the same hyperparameters of the baselines as our model.

\begin{table}[!t]
\begin{small}
\centering
\caption{ Comparison of recommendation performance in Movie-Music (\%). The best results are in \textbf{bold} and the second best ones are \underline{underlined}.}

% \vspace{0.5cm}

\begin{tabular}{p{1.95cm}|p{1.15cm} p{1.15cm}|p{1.15cm} p{1.15cm}}
\toprule
 & \multicolumn{4}{c}{Movie-Music (MM)} \\ \cline{2-5} 
\multirow{2}{*}{Methods} & \multicolumn{2}{c|}{Movie}  & \multicolumn{2}{c}{Music} \\ \cline{2-5} 
& MAE & F1\_Score  & MAE & F1\_Score \\ \midrule

MF \cite{koren2009matrix} & 20.94$\pm$2.54 & 74.97$\pm$4.50&  23.79$\pm$1.69 &  72.57$\pm$0.75  \\

NCF \cite{he2017neural}  & 19.01$\pm$0.09 & 88.93$\pm$0.05 & 15.25$\pm$3.23 & \underline{93.05$\pm$0.43} \\ \midrule

CMF  \cite{singh2008rel} & 20.23$\pm$1.97 & \underline{89.09$\pm$0.36} & \underline{11.66$\pm$2.35} & 92.45$\pm$0.36   \\

CoNET \cite{hu2018conet}& \underline{18.22$\pm$0.36} & 88.68$\pm$0.70 &  13.96$\pm$0.36 & 92.05$\pm$0.48 \\

 DDTCDR \cite{zhu2019dtcdr} &20.69$\pm$0.35& 74.84$\pm$1.74 &15.82$\pm$0.75 &  89.05$\pm$2.13  \\ \midrule

% Ours0 & 14.23$\pm$0.97 &90.69$\pm$0.22 &9.89$\pm$0.35 & 94.45$\pm$0.32  \\
Ours & \textbf{14.23$\pm$0.97} &\textbf{90.69$\pm$0.22} &\textbf{9.89$\pm$0.35} &\textbf{94.45$\pm$0.32}  \\
Improvement (\%) & 21.28 \% & 1.80 \% & 15.18 \% & 1.50\% \\
% Ours2 & 14.23$\pm$0.97 &90.69$\pm$0.22& 9.89$\pm$0.35 & 94.45$\pm$0.32  \\
% Ours2 & 7.07$\pm$0.22	& 14.54$\pm$0.17 &90.92$\pm$0.22& 4.36$\pm$0.14 &9.03$\pm$0.51 & 93.75$\pm$0.32  \\
\bottomrule
\end{tabular}
\label{tab:movie_music}
\end{small}
\end{table}

\begin{table}[!t]
\begin{small}
\centering
\caption{ Comparison of recommendation performance in Movie-Book(\%). The best results are in \textbf{bold} and the second best ones are \underline{underlined}.}
% \vspace{0.5cm}
\begin{tabular}{p{1.95cm}|p{1.15cm} p{1.15cm}|p{1.15cm} p{1.15cm}}
\toprule
\small
 & \multicolumn{4}{c}{Movie-Book (MB)} \\ \cline{2-5} 
\multirow{2}{*}{Methods} & \multicolumn{2}{c|}{Movie} & \multicolumn{2}{c}{Book}  \\ \cline{2-5} 
 & MAE & F1\_Score  & MAE & F1\_Score \\ \midrule

MF \cite{koren2009matrix} & 24.17$\pm$1.32 & 73.64$\pm$0.74 &  23.83$\pm$1.25 & 69.01$\pm$2.74 \\

NCF \cite{he2017neural} &18.80$\pm$0.54& 89.08$\pm$0.07 &18.86$\pm$0.52& 89.35$\pm$0.06 \\ \midrule

CMF \cite{singh2008rel} & \underline{14.53$\pm$1.51} & 89.32$\pm$0.04 &  \underline{13.22$\pm$0.78} &  89.07$\pm$0.22\\

CoNET \cite{hu2018conet} & 17.46$\pm$0.61 &  \underline{89.59$\pm$1.45}  & 17.18$\pm$0.59&  89.22$\pm$0.77\\

DDTCDR  \cite{zhu2019dtcdr} & 20.17$\pm$0.56 & 82.60$\pm$2.37  & 17.15$\pm$0.54 &  \underline{90.06$\pm$0.39} \\ \midrule

Ours  & \textbf{13.17$\pm$0.16} & \textbf{90.60$\pm$0.37}  & \textbf{13.01$\pm$0.14} & \textbf{90.80$\pm$0.22} \\
% Ours2  & 7.03$\pm$ 0.39 & 14.63$\pm$0.11 & 90.72$\pm$0.17  &  6.34$\pm$0.11& 13.39$\pm$0.34 &  90.52$\pm$0.08  \\
\small Improvement (\%) & 9.36 \% & 1.12 \% & 1.58 \% & 0.57\% \\
\bottomrule
\end{tabular}
\label{tab:movie_book}
\end{small}
\end{table}

 \begin{figure}[!t]
%	\vspace{0.5cm} 
	\centering
	\includegraphics[width=0.5\textwidth]{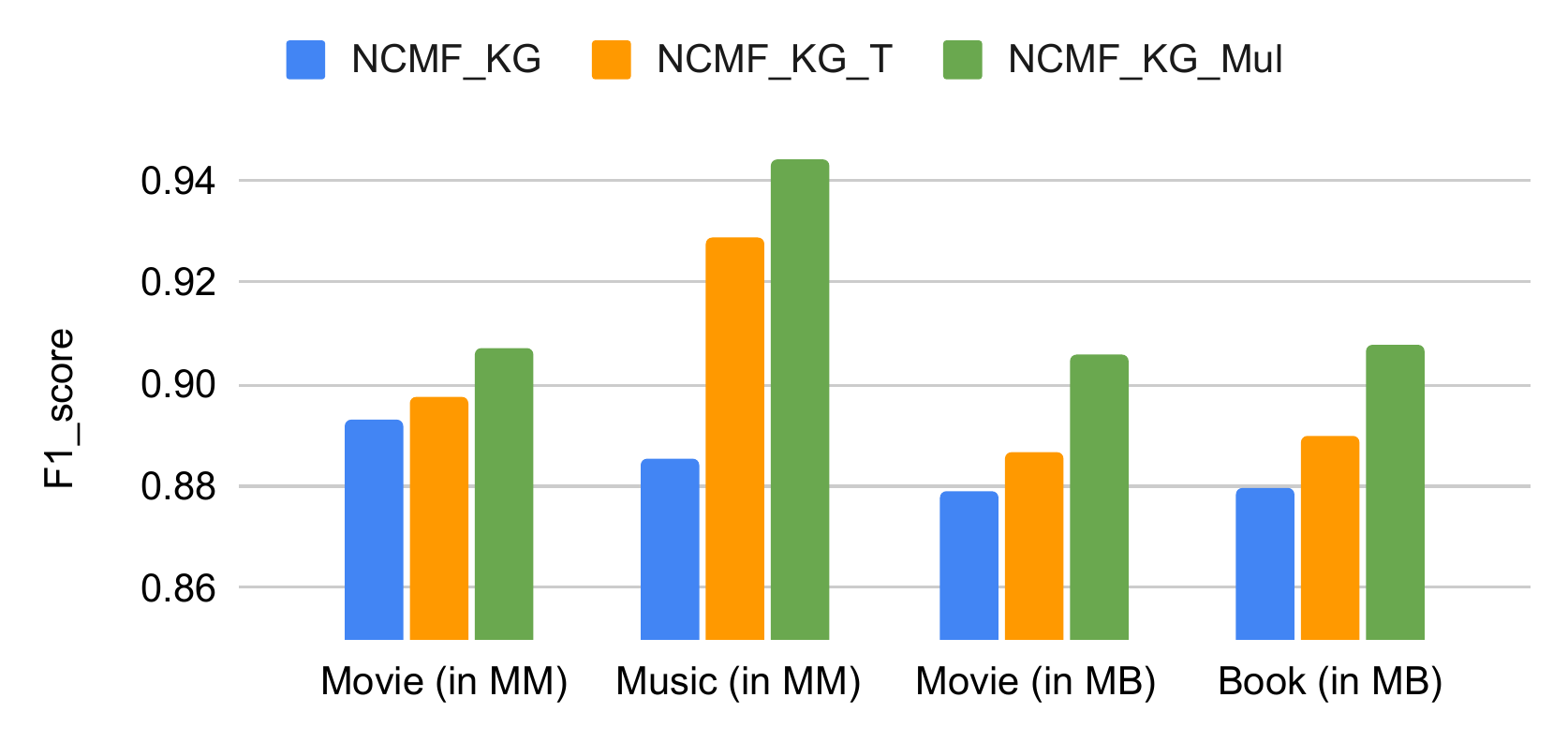}
	\caption{Different ways to incorporate KG information for CDR.}
	\label{fig:different_integration}
\end{figure}

\subsection{Overall Performance of CDR}
We have conducted experiments on two cross domain tasks, movie-music (MM) and movie-book (MB), and the corresponding results of our model and baselines are shown in Table \ref{tab:movie_music} and Table \ref{tab:movie_book}. We can see that our proposed model can consistently obtain the best performance across movie-music and movie-book recommendations in terms of MAE and F1\_{score}. In particular, our model improves over the strongest baselines $w.r.t.$ MAE by 21\%, 15.18\% in movie, music (Table \ref{tab:movie_music}) respectively, which justifies the effectiveness of our method in integrating items' KG information. If we compare between these two tasks, MM and MB, the improvement on music in MM is more remarkable compared to the performance in MB. Possible reasons are 1) the data is more sparse in the user-music interaction matrix, %compared to others, 
so leveraging KG information can greatly relieve the sparsity problem (we have verified this in the later experiments: Comparisions for cold-start item scenarios); 2) the extracted KG contains much useful information, especially for two closely related domains (movie and music both belong to multi-media datasets). Besides, CDR models (CMF,CoNet,DDTCDR) achieve better performance than SDR models (MF, NCF), indicating that utilizing extra information from other resources benefits the performance of recommendation.

\subsection{Different ways to incorporate KG}
We explore different ways to combine item embeddings learned from KG and user-item interaction matrices. NMF\_KG takes KG-level embeddings as input, then incorporates them with item embeddings learned from user-item interaction matrices via an aggregation method, e.g., concatenation.  NCMF\_KG\_T tries to refine item embeddings learned from KG with a one-layer MLP and concatenates with embeddings learned from the user-item interaction matrix. NCMF\_KG\_mul maximizes MI between the two types of representations to guarantee the highly correlated relationship. The results are shown in Fig. \ref{fig:different_integration}. Generally, refining the learned KG-level embeddings gets better performance than direct utilization. This is because in real-world KGs (e.g., Freebase) some noises are inevitably introduced in the process of automatically constructing large-scale KGs due to limited labour supervision \cite{Xie2018DoesWS,Jia2019TripleTM}. NCMF\_KG\_mul gets the best performance. The possible reason is that item embeddings  jointly learn from the user-item rating matrix and entity embeddings from KG, which contain both domain-general and domain-specific knowledge and the neural mutual information estimator can ensure their correlation. Such design is more suitable for the cross-domain recommendation task.

\subsection{Comparisons for cold-start item scenarios}
The KG a natural bridge for items from different domains, which can further alleviate the item cold-start problem in RS. 

\begin{figure}[!t]
\centering
  \begin{subfigure}[b]{0.48\linewidth}
	\centering
		\includegraphics[width=\textwidth]{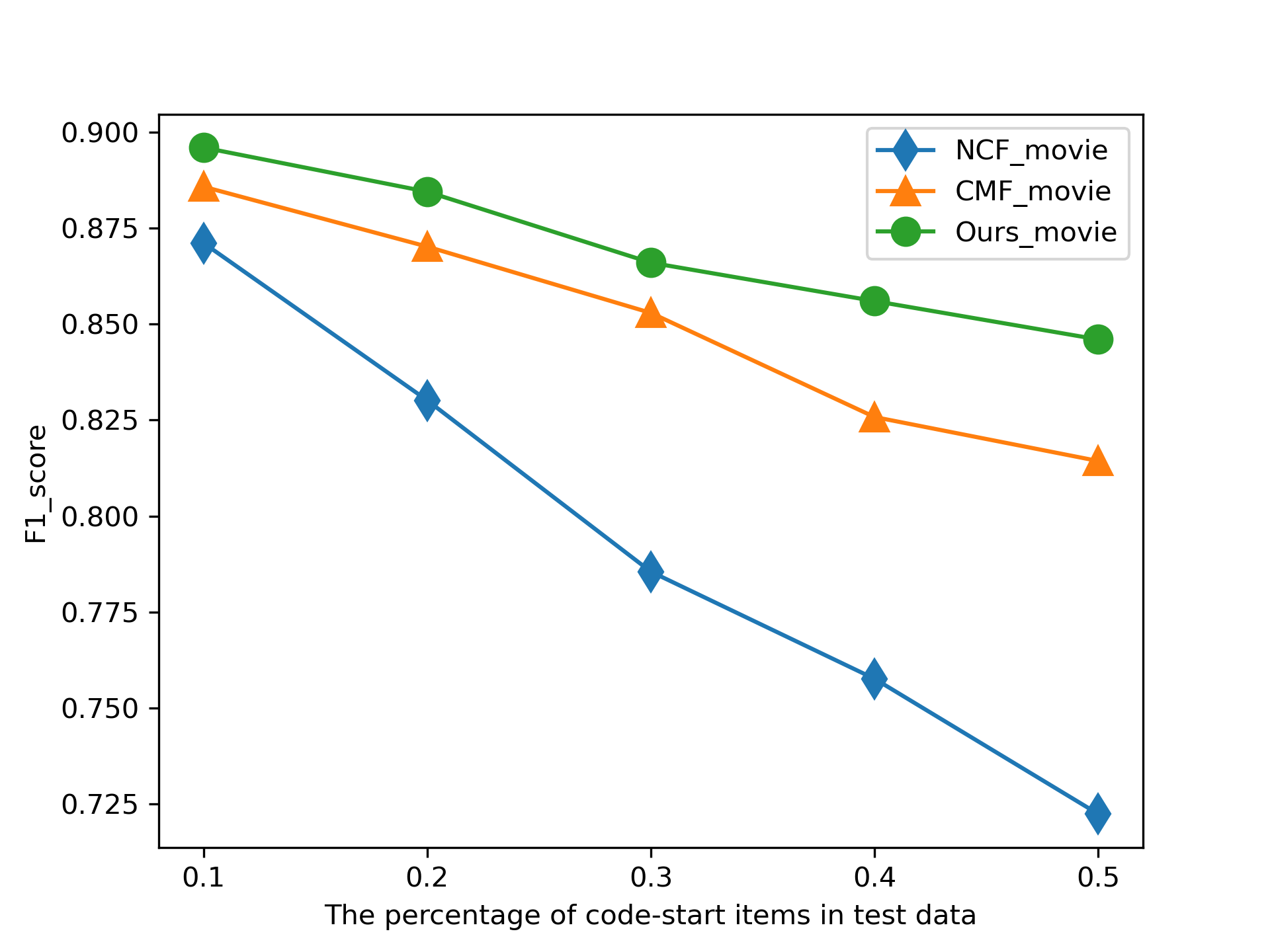}
	     \caption{Movie (in MM).}
		\label{fig:cold_movie}
	\end{subfigure}
	\begin{subfigure}[b]{0.48\linewidth}
		\includegraphics[width=\textwidth]{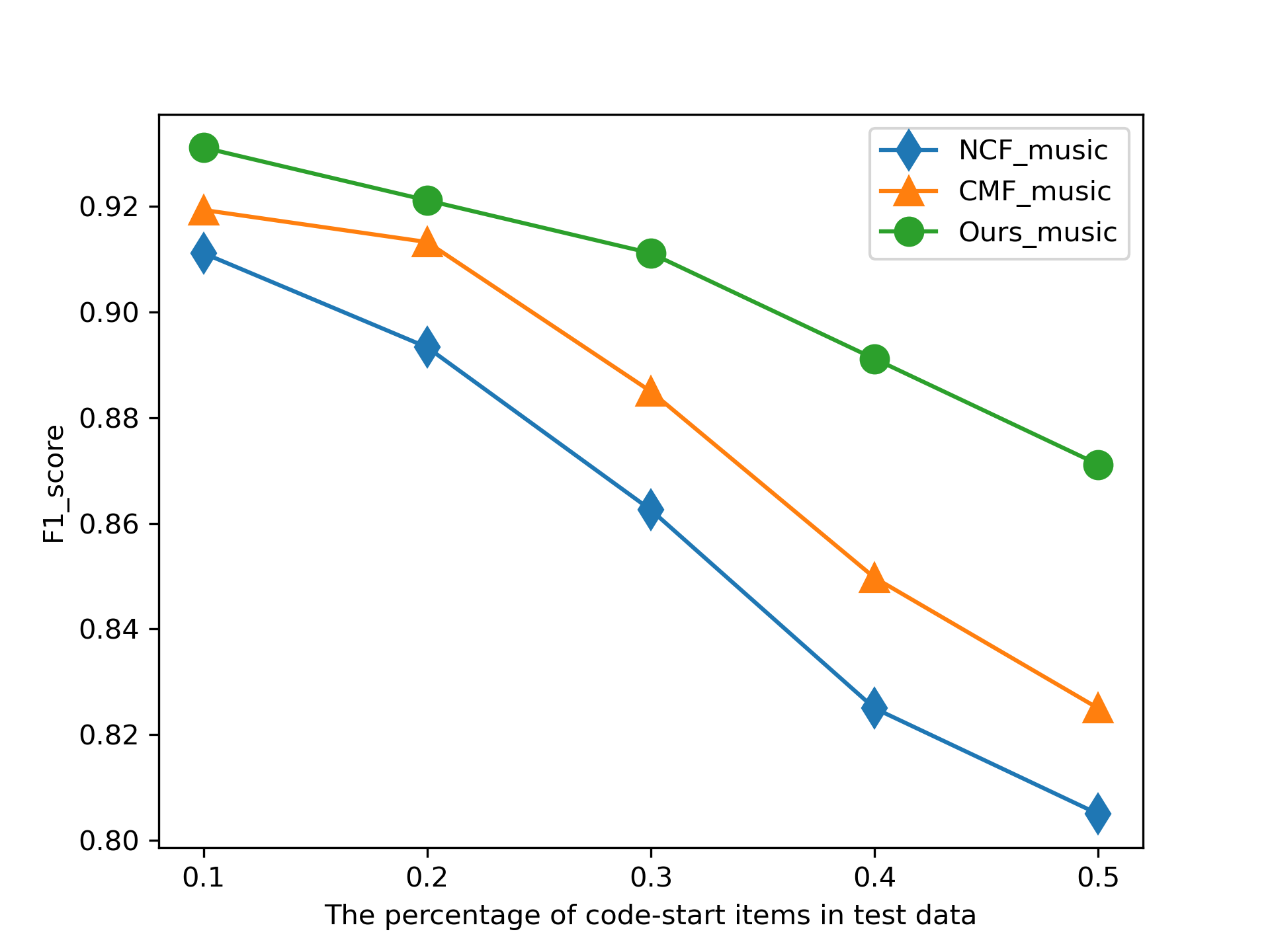}
		\caption{Music (in MM).}
		\label{fig:cold_music}
	\end{subfigure}
	\caption{Comparison of different models in cold-start items scenarios.}
	\label{fig:cold-start}
	\vspace{-0.5cm}
\end{figure}

To validate this, we compare our methods with NCF, CMF under the code-start scenario. We set up the cold-start environment by sampling a subset of items for testing which are unseen in the training data.
Results for cold-start items on movie-music datasets are shown in Fig. \ref{fig:cold-start}. NCF (the SDR model) is greatly influenced and gets the poorest performance, especially there are a large proportion new items. CMF (the CDR model) can leverage information from two domains, thus it can alleviate the cold-start problem in some extent. Our model goes further to learn representations for cold items from the KG, offering additional information beyond user-item interaction matrices.

\section{Conclusion}
In this paper, we constructed a new dataset \textit{AmazonKG4CDR}, the first in the filed linking KG information for cross-domain recommendation. Moreover, we proposed a KG-aware NeuCMF model to learn domain-specific and domain-general knowledge using graph autoencoding strategy to capture both adjacent and higher-order neighborhood information from KG. Our model unified item embeddings learned from user-item interaction matrices and KG with a neural collaborative filtering framework under a mutual information-based neural estimator. Through extensive experiments on real-world datasets, we demonstrated that KG-aware NeuCMF has achieved substantial gains over state-of-the-art baselines. For future work, we will explore the explainability of cross-domain recommendation.

% This work used a two-step learning framework and adopted the strategy of multi-task learning, a future direction could be  to jointly learn the recommendation task with the guidance of the KG-related task 

% In this paper, we constructed a new dataset \textit{AmazonKG4CDR}, the first in the filed linking KG information for cross-domain recommendations. Further, we proposed KG-aware NeuCMF model that naturally incorporates knowledge graph information in the cross-domain recommendation. Our model can learn domain-specific and domain-general knowledge by by utilizing graph autoencoding strategy, which can capture both adjacent and higher-order neighborhood information from KG. KG-aware NeuCMF unified item embeddings learned from the user-item interaction matrix and KG with a neural collaborative filtering framework under a mutual information-based neural estimator. Through extensive experiments on real-world datasets, we demonstrate that KG-aware NeuCMF achieves substantial gains over state-of-the-art baselines. In this work, we used a two-step learning framework and adopted the strategy of multi-task learning, to jointly learn the recommendation task with the guidance of the KG-related task is an important direction of our future work.

\bibliographystyle{ACM-Reference-Format}
\balance
\bibliography{oldbib.bib}
\end{document}